# Comparing fast imaging techniques for individual pulse imaging by Cherenkov *in vivo* from electron FLASH irradiation

**Mahbubur Rahman**[1*]**, M. Ramish Ashraf**[1,2]**, Rongxiao Zhang**[1,3,4]**, Xu Cao**[1,5]**, David J. Gladstone**[1,3,4]**, Lesley A. Jarvis**[3,4]**, P. Jack Hoopes**[1,3,4,6]**, Brian W. Pogue**[1,4,6,7]**, Petr Bruza**[1*]

[1] Thayer School of Engineering, Dartmouth College, Hanover NH 03755, US

[2] Department of Radiation Oncology, Stanford University, Stanford, CA 94305

[3] Department of Medicine, Radiation Oncology, Geisel School of Medicine, Dartmouth College Hanover NH 03755 USA

[4] Norris Cotton Cancer Center, Dartmouth-Hitchcock Medical Center, Lebanon, NH 03756 USA

[5] School of Life Science and Technology, Xidian University, Xi'an, China

[6] Department of Surgery, Geisel School of Medicine, Dartmouth College, Hanover NH 03755 USA

[7] Department of Medical Physics, University of Wisconsin, Madison WI 53705 USA

[*]Correspondence to: Mahbubur.Rahman.TH@dartmouth.edu or Petr.Bruza@dartmouth.edu

**Abstract:**

**Objective:** In this study, a fast-imaging technique was developed for the first *in vivo* Cherenkov emission imaging from an ultra-high dose rate (UHDR) electron beam source at single pulse (360 Hz) submillimeter resolution.

**Approach:** A CMOS camera, gated to the UHDR LINAC, imaged the Cherenkov emission profiles pulse by pulse passively during the irradiation of mice on their limbs and intestinal region. The utility of an intensifier was investigated for its effect on image quality including signal to noise and spatial resolution. Pulse by pulse variability in Cherenkov emission profile were quantified spatially and temporally.

**Main results:** An intensifier improved the emission profile's signal to noise ratio from 15 to 280, with reduced spatial resolution. The profile extended beyond of the treatment field due to the lateral scattering of the electrons in tissue and its optical properties. The CMOS camera with an intensifier detected the changes in Cherenkov emission profile during expiration and inspiration of the respiration cycle for the mice to be ~3 mm.

**Significance:** This fast-imaging technique can be utilized for in vivo intrafraction monitoring of FLASH patient treatments at single pulse resolution. It can display delivery differences during respiration, and variability in the delivered treatment's surface profile, which may perturb from the intended UHDR treatment more for pencil beam scanning systems. The technique may leverage Cherenkov emission surface profile to gate the treatment delivery via respiratory gating systems under FLASH conditions.

1. Introduction

Since first described by Hornsey et al in 1966[1], ultra-high dose rate (UHDR) irradiation for investigation of the FLASH effect has seen a resurgence led by Favaudon et al in 2014[2]. The studies demonstrated that UHDR (>40Gy/s) treatments reduced normal tissue toxicity with equivalent tumor control when compared to conventional treatment delivery (~0.1 Gy/s). Since then there were a plethora of experiments and studies on various biological endpoints[3–8] further confirming the differential response from FLASH treatments. There have been clinical experiences with FLASH irradiation as the first human patient and canines were treated with an experimentally dedicated and converted clinical LINAC with UHDR electron beams, respectively[9,10]. Nonetheless, some studies indicated the lack of FLASH effect from UHDR beams[11,12] which may due to the biological endpoints and models considered. It may also be due to dosimetry considerations such as the temporal structure of the conventional and UHDR beam. Studies have shown that there are dosimetry uncertainties with preclinical radiobiological studies that can affect the reproducibility and translation of new technology or modality[13,14].

Cherenkov emission imaging provides a potential solution as it characterized treatment irradiators and utilized as a dosimetry tool for preclinical and clinical studies. Due to its isotropic emission in tissue and linear relationship with dose[15], it is utilized for imaging both humans and animals. Cameras imaged Cherenkov light from many irradiation sources including radioactive nuclides[16–18], medical isotopes[19], external beam radiotherapy[15], and proton beams[20]. The emission source quantified dose[21,22], verified patient positioning[23,24], confirmed match line for multiple field delivery[25]. Under UHDR conditions, Favaudon et al demonstrated Cherenkov emission's near instantaneous production in a medium can temporally resolve dose of their UHDR electron beam source[26]. Rahman et al demonstrated using an intensified CMOS camera and a water phantom, an UHDR electron beam can be characterized via Cherenkov emission at millimeter and single pulse resolution (60 fps)[27].

This study investigates a fast-imaging technique of the first *in vivo* Cherenkov emission from an UHDR irradiator using an intensified CMOS camera. The 10 MeV electron beam[28] source was from a modified clinical LINAC. Mice were imaged during FLASH treatments at single pulse submillimeter resolution and 360 frames per second. The effects of an intensifier (photocathode) on the image quality were considered concerning signal to noise ratio and spatial resolution. The Cherenkov emission temporal and spatial profiles of the treatments were provided. The potential utility of the technology during FLASH irradiation as a dosimetry tool were discussed which included monitoring respiration during treatment and confirming dose rate distribution of individual pulses *in vivo*.

## 2. Materials and Methods

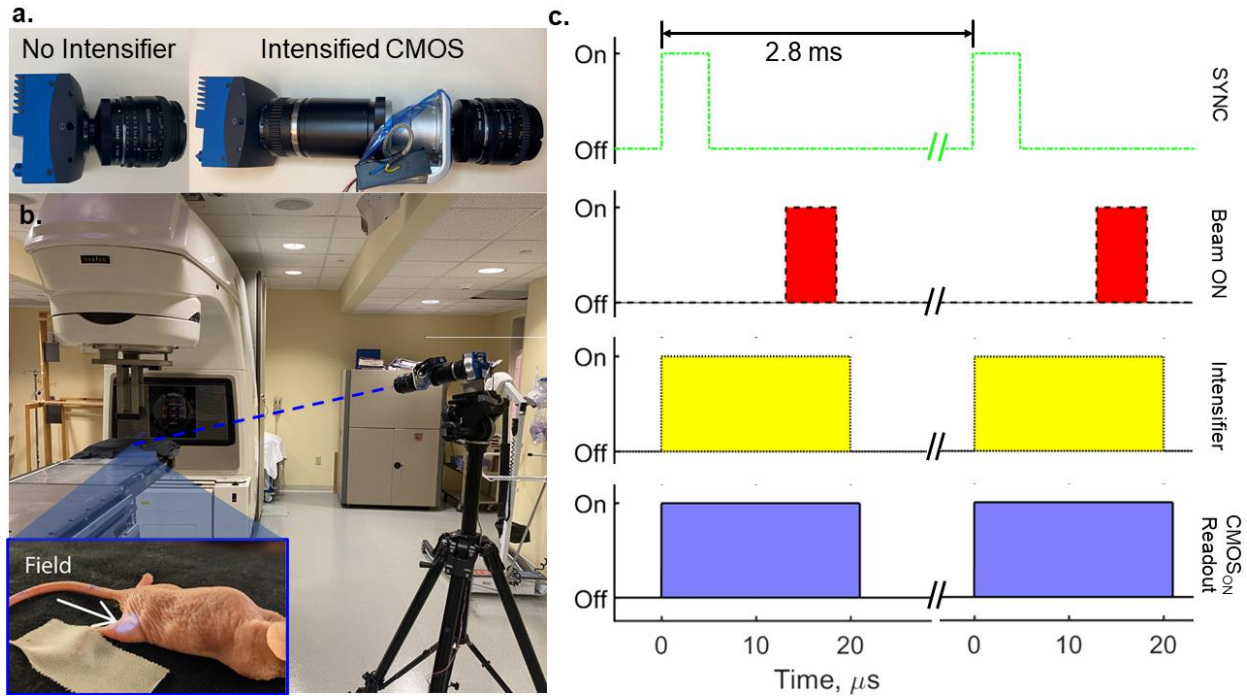

**Figure 1. a.** Experimental setup of Imaging acquisition of mice irradiated with electron FLASH beam. **b.** Sketched timing Diagram of Camera acquisition with respect to SYNC signal from LINAC and beam on time. The acquisitions occurred at 360 Hz to match the repetition rate of the LINAC. CMOS and the Intensifier were triggered off the rising edge of the SYNC signal. Images were captured with and without a red sensitive intensifier and the sensitivity spectrum presented in Rahman et al. 2021[29].

### 2.1. FLASH Treatment Delivery and LINAC Control System

Mice were treated as part of ongoing studies on investigating tumor and normal tissue outcome from delivery of >40 Gy/s ultra-high dose rate (UHDR) beams i.e. the FLASH effect. All animals were cared for and handled in accordance with National Institutes of Health guidelines for the care and use of experimental animals and study protocol. The nude mice were kept under general anesthesia administered as inhaled isoflurane at 1.5% via flowing air through a nose cone during all treatment and measurement procedures. A heating pad was used to maintain the normal physiological body temperature.

A Varian Clinac 2100 C/D (Palo Alto, CA) was modified to deliver UHDR 10 MeV electron beams at the isocenter[28] and the surface of the mice's' limbs or the intestinal region were aligned to the isocenter. The UHDR electron beam with a 1 cm cutout and an average output of 0.66 Gy/pulse delivered treatment to the limbs of the mice. The beam with a 1.5 cm cutout and output of 0.72 Gy/pulse delivered treatment to the intestinal region of a mouse. The LINAC delivered pulses at 360 Hz repetition rate for a mean dose rate of ~240 Gy/s and ~260 Gy/s, respectively. The number of pulses delivered to mice were determined based on the EBT-XD Gafchromic film measured output and the prescribed dose. The pulses delivered were controlled and gated using a coincidence based scattered radiation detector (DoseOptics LLC, NH), an Arduino Mega 2560 (Arduino LLC, MA) control circuit, and gating switchbox (Varian Inc, CA). The control circuit counted the pulses delivered measured by the detector and sent signal to the gating switchbox, which halted the beam using the MLC hold-off signal once the prescribed pulses were delivered. The dose delivered was confirmed using either EBT-XD film placed on the applicator (and conversion factor for dose at the isocenter) or below the mice for the ones treated on the limb.

## 2.2. Camera Acquisition

The Cherenkov emission profiles were measured pulse by pulse passively during the irradiation of the mice as shown in **Figure 1a**. The Quad Channel CXP-12 GigaSens camera (Concurrent EDA, PA) was positioned to image the entire treatment field. As indicated by **Figure 1b.** the camera's CMOS sensor and photocathode intensifier (Photonis Scientific Inc., MA) was gated to the SYNC signal from the LINAC. The SYNC signal preceded the delivery of each beam by ~13µs (measured by the coincidence based scattered radiation detector) both with a pulse width of ~5µs. Thus, the CMOS sensor had an exposure and readout time of 21µs, capturing the entirety of each pulse. For comparison of an intensifier's effect on image quality, a mouse was imaged with and without the intensifier which was also triggered from the SYNC signal with an on time of 20μs. The camera acquired images at 360 frames per second to match the repetition rate of LINAC.

## 2.3. Image Processing and Emission Profile Characteristics

The images were processed to ensure quality of the Cherenkov emission profiles. A dark field image was subtracted from each frame and corrected by a flat field image accounting for any variable response of each CMOS pixel. The frames were 3D median filtered (3×3×3 voxels) to remove stray radiation signal and example frames are shown in **Figure 2a**. Pulse by pulse output were determined by summing the Cherenkov light in each frame, as shown in **Figure 2b**. The image quality was quantified based on signal to noise ratio (SNR) with the signal measured on a region of interest (ROI) on the Cherenkov emission profile and the noise measured on the skin of the mice outside of the treated field. The spatial resolution comparison was described qualitatively. The change in Cherenkov emission profile during respiration of the mouse treated on the intestinal region was from the difference in the intensity profiles during inspiration and expiration.

## 3. Results

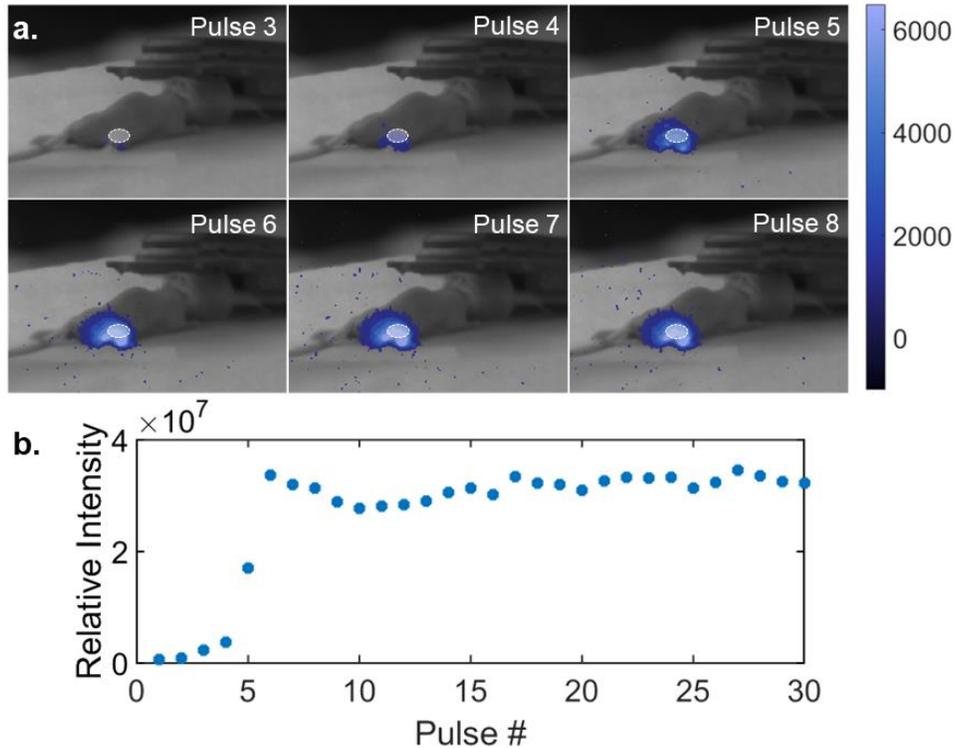

**Figure 2. a.** Example Cherenkov emission profiles from single pulse deliveries to the flank of mouse for a 1 cm diameter field overlayed on a white light background image of a mouse. **b.** Cumulative intensity per frame and pulse.

As evident from Figure 2, the camera was able to image at single pulses and 360 Hz providing spatial Cherenkov emission profiles at submillimeter resolution. There is a ramp up in the dose per pulse delivered for the first 4-6 pulses from the linear accelerator as shown in the representative frames and cumulative intensity per pulse, consistent with a previous study[28]. The representative frames in figure 2a were from pulses 3-8 showing the Cherenkov profile changing from pulse to pulse. Pulse 3 illustrated that the camera imaged little Cherenkov emission but pulse 6-8 demonstrated a consistent profile in Cherenkov emission spatially and via intensity, indicative of the beam stabilizing. While the field size was 1 cm in diameter, the spatial distribution of Cherenkov emission was measured to be beyond the 1 cm in diameter when considering full width half max (FWHM).

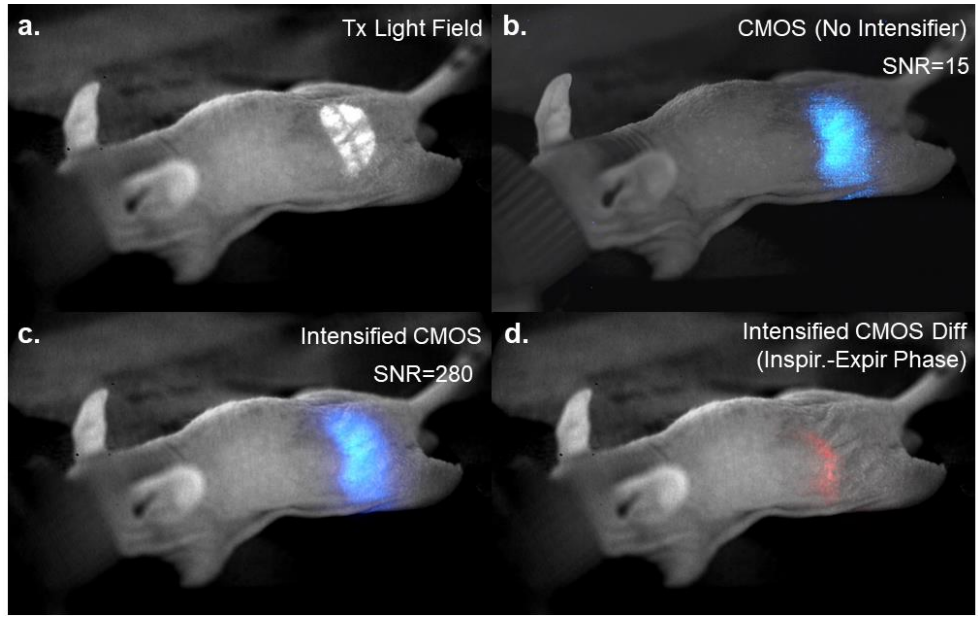

**Figure 3. a.** Light field of mouse treated with a 2 cm diameter field **b.** Cherenkov emission profile of a single pulse with no intensifier **c.** Cherenkov emission profile of a single pulse with an intensifier **d.** Difference in Cherenkov emission profile while the mouse breathed during delivery

While quantifying the Cherenkov emission image quality from imaging irradiation of a mouse's intestinal region, the changing emission profile could be seen due to it breathing as shown in figure 3. The measured emission profile without the intensifier had a SNR of 15, and with the intensifier the SNR was 280. There was a reduced spatial resolution with the intensifier as the images were blurrier. Like the mouse treated in figure 2, the Cherenkov emission spatial profile's FWHM in figure 3 seems to spread slightly beyond the treated field. With the intensifier the changes in Cherenkov emission profile were seen while the mouse was free breathing. Figure 3b. shows the difference during expiration and inspiration of approximately 3 mm.

## 4. Discussion:

This study demonstrates the first *in vivo* imaging of Cherenkov emission that provided spatiotemporal characteristics of treatment delivery from an UHDR irradiation source. The camera imaged the treatment delivery pulse to pulse, at 360 Hz capturing dose rates up to ~ 260

Gy/s. Figure 2 shows the emission profiles and output at single pulses. From figure 2a, and figure 3, it is evident that the spatial distribution of the Cherenkov emission profile is non-uniform with the greatest intensity at the center of the beam, where it achieved ~240 Gy/s, and ~260 Gy/s, respectively. The Cherenkov emission profile seemed to be beyond the treatment field for both 1cm and 1.5 cm diameter treated mice which may be indicative of lateral scattering of the electrons that irradiated the body part and was true dose outside of the treatment field. However, the larger profile may also be due to the optical properties of the naked mice allowing for subsurface imaging of the Cherenkov profile, projecting a larger profile than the treatment field on the camera. Nonetheless, Cherenkov emission in tissue was found to be isotropic[29], suggesting the intensity profile per frame can be correlated to a spatial distribution in the dose rate. The dose rate distribution is further complicated by the ramp up in the LINAC during the first ~4-6 pulses as shown in figure 2b. The single pulse Cherenkov imaging of these mice highlights that some portions of the delivery, whether penumbra region of the treatment field or certain times or pulses of the delivery, the dose rate distribution may be below the UHDR threshold of 40 Gy/s.

The single pulses in vivo Cherenkov imaging also illustrates a potential utility of this camera technology for FLASH treatments and builds on prior studies that utilized Cherenkov emission for dosimetry. Zlateva et al demonstrated Cherenkov emission can be utilized to characterize the spatial dose profiles of electron beam, requiring corrections for the depth and energy dependency[30,31]. Mitchell et al demonstrated that Cherenkov luminescence imaging can provide spatial distribution of the β emitting radionuclides or electron in vivo[16]. Under FLASH conditions, Favaudon et al demonstrated a probe that measure Cherenkov emission can provide time resolved dosimetry of UHDR pulsed electron beam[26]. In this study the Cherenkov emission imaged provided a spatiotemporal profile of mice, that can potentially lead to estimation of dose distribution, mean dose rate distribution, and dose per pulse (i.e. instantaneous dose rate) distribution. The dose rate distribution is of particular importance to FLASH studies as this can inform researchers on the treated tissue that received UHDR or sub-UHDR irradiation. It may help elucidate why certain UHDR experiments exhibited the FLASH effect when comparing UHDR vs conventional beams[11,32], while other studies showed no significant difference[12].

Utilizing a camera to image Cherenkov emission from mice irradiated with UHDR beams also benefits from delivery of dose at least 3 orders of magnitude intense than conventional beams for each individual pulse. Since the LINAC is delivering ~0.7 Gy/pulse, the camera achieved an SNR of 15 even without an intensifier as shown in Figure 3c. Nonetheless, the intensifier improved the SNR to 280, which suggests an improved dynamic range and allowed the camera to resolve the Cherenkov emission profile during the ramp up pulses shown in figure 2. The improved image quality may have contributed to resolving the difference in Cherenkov emission profiles during respiration of the mouse in figure 3b. However, there is reduced spatial resolution utilizing the intensifier that may be attributed to the photocathode.

As FLASH approaches clinical translation with the first patient already treated with an electron UDHR beam[9], imaging of the Cherenkov emission irradiated with an UHDR beams may highlight potential impact of FLASH irradiation on dose delivery consistency to patients. This tool can provide independent documentation and verification of dose per pulse output and spatial profile consistency. Figure 3c indicated that for free beathing treatment, even in a delivery of less than a second, the respiration can contribute to variability in dose delivery per pulse. This may have implication on patients such as ones treated on the breast where it has been shown that inspiration reduces dose to normal structures such as heart and liver[33]. While in the treated mouse, the observed motion was ~3 mm, for patients the motion amplitude can be as much as 1 to 3 cm[34]. The breathing motions becomes a bigger concern from delivery with a UHDR proton pencil beam scanning (PBS) system because scanning requires an elongated time of delivery in comparison to the step and shoot method of the electron beams and passive scatting FLASH proton beams[35]. Bruza et al[20] demonstrated Cherenkov emission can be imaged from proton pencil beam scanning systems, thus the technology presented in this study may potentially be utilized to passively monitor the beam profiles of UHDR PBS treated patients during respiration as well. With a higher frame rate, and assuming comparable signal to noise ratio the intensified fast camera may be utilized for imaging UHDR PBS treatments with dose rates of ~100 Gy/s[36].

## 5. Conclusion:

In this study, single pulse in vivo Cherenkov profile imaging of UHDR electron beams was demonstrated at 360 frames per second. CMOS imaging with an intensifier improved the signal to noise ratio with reduced spatial resolution in comparison to imaging without an intensifier. The Cherenkov emission profile extended beyond of the treatment field due to the lateral scattering of the electrons in tissue and its optical properties. This fast-imaging technique can potentially provide dose profiles during FLASH irradiation at single pulse resolution and can monitor delivery differences caused during respiration, which may perturb the intended treatment particularly during UHDR delivery from proton PBS systems. The fast-imaging technique may be utilized to document the treatment delivery as an independent quality control and patient monitoring tool.

**Acknowledgements**

This work was supported by the Norris Cotton Cancer Center seed funding through core grant P30 CA023108 and through seed funding from the Thayer School of Engineering, as well as support from grant R01 EB024498 and R43 CA268466-01.

**Conflict of Interest Statement:**

Mahbubur Rahman is a former employee of DoseOptics LLC. Dr. Zhang has a patent 10201718 issued, and a patent 15160576 issued. Dr. Pogue reports personal fees and other from DoseOptics LLC, outside the submitted work. Dr. Bruza reports non-financial support from DoseOptics LLC, during the conduct of the study; personal fees and non-financial support from DoseOptics LLC, outside the submitted work. In addition, Dr. Bruza has a patent 62/967,302 pending, a patent 62/873,155 pending, a patent PCT/US19/14242 pending, and a patent PCT/US19/19135 pending.